# Who Watches the Watchmen?

# A Review of Subjective Approaches for Sybil-resistance in Proof of Personhood Protocols

Divya Siddarth[1], Sergey Ivliev[2], Santiago Siri[3], Paula Berman[4*]

**\* Correspondence:**
Paula Berman
paula@democracy.earth

**Abstract.** Most self-sovereign identity systems consist of strictly objective claims, cryptographically signed by trusted third party attestors. Lacking protocols in place to account for subjectivity, these systems do not form new sources of legitimacy that can address the central question concerning identity authentication: "Who verifies the verifier?". Instead, the legitimacy of claims is derived from traditional centralized institutions such as national ID issuers and KYC providers. This architecture has been employed, in part, to safeguard protocols from a vulnerability previously thought to be impossible to address in peer-to-peer identity systems: the Sybil attack, which refers to the abuse of a digital network by creating many illegitimate virtual personas. Inspired by the progress in cryptocurrencies and blockchain technology, there has recently been a surge in networked protocols that make use of subjective inputs such as voting, vouching, and interpreting, to arrive at a decentralized and sybil-resistant consensus for identity. In doing so, these projects illustrate that the best technologies do not abstract away subjectivity but instead embrace it as a necessity, and strength. In this review, we will outline the approaches of these new and natively digital sources of authentication - their attributes, methodologies strengths, and weaknesses - and sketch out possible directions for future developments.

## 1    Introduction

For blockchain networks to move from strictly providing financial services into enabling social and political applications, decentralized protocols for verifying unique human identities must be devised. Lacking reliable means to do so, currently most blockchain governance practices validate membership by employing Proof of Stake (requiring ownership of a given cryptocurrency) or Proof of Work (requiring ownership and use of mining hardware). These resource-based membership systems have rendered most crypto-governance practices into plutocracies, with a few powerful players able to control choices according to their own interests (De Filippi 2019). Clearly, this is antithetical to democratic principles. If blockchains are to become a significant public infrastructure, particularly in the space of civic engagement, then Proof of Work's "one-CPU-one-vote" or Proof of

---

[1] RadicalXChange Foundation
[2] Perm State University
[3] Democracy Earth Foundation
[4] Democracy Earth Foundation

Stake's "one-dollar-one-vote" systems will not suffice: in order to enable democratic governance, protocols that signal unique human identities to enable "one-person-one-vote" systems must be created.

At the center of this question of identity is the Sybil attack problem. Previously thought to be impossible to address in distributed identity systems, Sybil attacks describe the abuse of a digital network by creating many illegitimate virtual personas (Swathi, Modi, and Patel 2019; Ford 2012; Douceur 2002). When applied to the identity space this challenge has also been defined as the "unique-human" (or, more realistically, "semi-unique human") problem[5][6]. However, there has recently been a surge in networked protocols that make use of subjective inputs - such as voting, vouching, and interpreting - to arrive at a decentralized and sybil-resistant consensus for identity. Driven by the goal of creating a digital layer for humanity, with on-chain and off-chain governance structures (De Filippi and McMullen 2018), where peers are able to vote, organize and transact freely, recent projects have been contributing significant learnings to the domain. The present article aims to provide a cohesive review of these learnings, and of the projects currently being built to solve for this type of identity protocol.

We begin by briefly reviewing previous approaches to natively digital identity. We cover both decentralized approaches, with the resource-based mechanisms of Proof of Work and Proof of Stake; and centralized approaches, with private internet platforms that function as credential providers. In Section 3 we develop a formal definition of Proof of Personhood as a new class of identity protocols aimed at providing a solution for the Sybil Attack via subjective approaches. We then outline the established primary properties of these protocols as well as their applications. In Section 4 we develop a taxonomy of approaches that have previously been employed to address the seminal question,"How can we distinguish a human from a machine?", and serve as the underlying primitives of Proof of Personhood protocols. In Section 5 we outline the leading implementations in the space - their attributes, methodologies, strengths, and weaknesses - and sketch out possible directions for future developments. In Section 6 we discuss these projects holistically, in the context of strengths, weaknesses, and tradeoffs as conceptualized in the space. We then outline the research gaps evident in each protocol, with the hope of providing a path forward to addressing and solving these issues.

The below information is compiled through a review of the academic literature in the space, as well as a compilation of secondary research on each of the approaches discussed, many of which have not yet been studied by the academy. We note that, while the presented protocols are by no means an exhaustive summary of all existing implementations, they do provide a diverse and relevant base from which to understand the state of current work, and to identify relevant tradeoffs and gaps in the space.

## 2     Previous Approaches to Identity Verification

First, we briefly review previous approaches to identity, both decentralized and centralized. In the blockchain space, this begins with Bitcoin's Sybil-protection mechanism, coined Proof of Work. This solution employs a resource-based membership model by proposing a challenge to nodes that requires computational work. The CPU that solves this challenge first obtains the right to add the next block of transactions to the chain, and also wins a reward in Bitcoin. If the other CPUs computationally agree with the validity of the event, they add that block to their own chains, and turn towards solving the challenge for the next block. The majority decision is represented by the longest chain, which has the greatest computational resources effort invested in it. Thus, Proof of Work is a "one-CPU-one-vote" system (Nakamoto 2009), and therefore centered around machine attributes, rather than subjective, human-centered inputs. Similarly, with governance practices relying on Proof of Stake —the main Sybil-protection alternative to Proof of Work— membership is established by the ownership of an asset, in this case, financial stake.

---

[5] 'Hard Problems in Cryptocurrency: Five Years Later', Vitalik Buterin (2019)
[6] 'Problems', Vitalik Buterin (2014)

The lack of a robust notion of personhood that could sustain a democratic governance model for blockchain networks has led to the development of plutocracies (De Filippi, Mannan, and Reijers 2020; De Filippi 2019): voting power is always relative to stake ownership (clearly so in the Proof of Stake case, and the problem remains even if stake consists of Proof of Work mining rigs). This results in a fat-tailed distribution of voting power in those systems, which reflects the Pareto distribution of wealth in society and financial markets (Klass et al. 2006; Benhabib, Bisin, and Zhu 2014).

The relevance of formalizing identities through natively digital proofs can also be inferred from centralized networks: major Internet platforms such as Facebook, Twitter, and Google established themselves partially by achieving a sufficient level of consensus over their identity credentials, thus creating a trust layer on top of which a myriad of social applications could be built. This has in turn facilitated the emergence and spread of multiple borderless political and social movements, as demonstrated by the role of social media in both national and international politics over the past few years (Tufekci 2017; Bruns et al. 2015; Bennett 2012).The only former alternative able to reach such widespread use pertained to nation-states: passports, licenses, and national ID cards. The creation of a global identification system outside of the strict control of nation-states has accelerated communication and knowledge creation, forming a networked social infrastructure that has allowed for a new kind of participative politics.

However, there are major vulnerabilities with this system, most pertinently i) privacy concerns and data misuse, and ii) the risk of creating exclusions to the system, with significant adverse social and political effects. The underlying architecture and ownership structure of these current centralized protocols exposes society to surveillance, political manipulation, and data theft; this is particularly relevant in our current global environment, marked by receding democratic freedoms and rising digital authoritarianism (Freedom House 2020; 2018).

It is also important to note that officially-recognized forms of ID are problematic for an estimated 1.1 billion people around the globe (Vyjayanti Desai 2017). Therefore, there is a clear need to create a protocol for identity consensus that can operate outside of centralized structures, whether they be nation-states or centralized and privately-owned platforms, while enabling the governance of blockchain networks to prevent the concentration of power and influence present in current Proof of Stake or Proof of Work systems. The efforts concerning the creation of a distributed and human-centered protocol coalesce into a third denomination: Proof of Personhood (Borge et al. 2017).

3.   **Proof of Personhood Protocols**

Research in the Proof of Personhood (henceforth PoP) ecosystem aims to extend and improve upon Proof of Work and Proof of Stake approaches, by focusing on methods capable of creating an analogous decentralized protocol to enable one-person-one-vote systems over blockchain networks. In order to lead to a sybil-resistant consensus for human identification, such a system needs to ensure that every identity within their domain is i) *unique*, so that no two people should have the same identifier, and ii) *singular*, so that one person should not be able to obtain more than one identifier (Wang and De Filippi 2020). The different protocols reviewed here aim to achieve sybil resistance while also maintaining self-sovereignty (anybody can create and control an identity without the involvement of a centralized third party) and being privacy-preserving (one can acquire and utilize an identifier without revealing personally identifying information in the process). Those three requirements, sybil resistance, self-sovereignty and privacy-preservation, compose the "Decentralized Identity Trilemma"[8]. Proof of Personhood approaches aim to achieve those three requirements, to different degrees, by establishing the following:

- *Subjective substrate.* Some form of "human entropy" that can act as a substitute for the computational work employed by the Proof of Work protocol, or the financial stake employed by the Proof of Stake protocol. This substrate can be expressed in the form of voting, interpreting, being present in a specific place (physical space or cyberspace) and time or interacting with others. Typically, the kind of substrate provided needs to be easy for humans to produce, but difficult for Artificial Intelligence to replicate, thus diminishing the ability of computer-generated false identities to take over the protocol. Additionally, this substrate needs to be relatively easy for humans to produce once, but relatively difficult for humans to produce two times or more, thus placating the ability of human-generated false identities to take over the protocol. A salient feature of these substrates is that they will typically involve minimal to zero personally identifiable information, thus preserving the privacy of authenticated individuals.

- *Objective incentive.* An incentive for nodes to join the network and continuously maintain its legitimacy. Ideally, this incentive needs to be strong enough to ensure that it is more valuable to be a part of the network as a legitimate entity, than selling one's membership as a Mechanical Turk. With the exception of Upala, all of the protocols described in this report employ or aim to employ some form of Universal Basic Income in cryptocurrency, associated with the protocol and distributed equitably to all members. This incentive can serve as a way to employ a system of behavioural economics, where one loses currency through misbehavior (by somehow attacking the legitimacy of the protocol), or earns more by behaving in ways that make the protocol stronger. Additionally, there may be other incentives, such as the desire for partial privacy or full anonymity in online spaces and transactions.

## 3.1 Primary Properties

We further outline the following desired primary properties of the Proof of Personhood protocols, consolidated from the literature, that allow for comparison:

- *Decentralization.* Decentralization is a multifaceted measure, analyzing how many independent parties have effective control over various components of the distributed system (Srinivasan 2017). In the case of PoP, the components of interest are the identity registry, graph of connections and vouching links, software code and releases, operation systems, blockchains and hardware. Sybil-resistant identity systems must have controls or incentives that would prevent control by bad actors, either through collusion or through purchasing verified identities en masse. Decentralization minimizes trust by eliminating third parties, and thus also maximizes collusion resistance. It is particularly important to identify who has permissions to write to the registry of identities: that is, whether the registry is permissionless or instead permissioned and controlled by an organization or a consortium. If the latter is the case, the protocol is not decentralized as the trust on a registry manager is required, even if the identity information is stored on a decentralized ledger.

---

[7] ['Not a Sybil'](#): Exploring the Path to Non-Dystopian Approaches to Digital Personhood, Aleeza Howitt, Daniel Burnet, et. al, 2019
[8] ['Decentralized Identity Trilemma'](#), Maciek, 2019.

- *Privacy preservation*. We analyze levels of privacy preservation through measures of anonymity, pseudonymity, unlinkability, unobservability, and plausible deniability (Beckers and Heisel 2012). A brief outline: Anonymity means that individuals are not identifiable. Pseudonymity denotes the usage of an identifier rather than a real name. Unlinkability refers to the fact that attackers cannot determine linkages between items such as transactions and addresses. Unobservability implies the anonymity of the persons involved in the transactions. The final aspect is plausible deniability: the ability to convincingly deny the possession of a certain identity and impossibility for authorities to prove the opposite.

- *Scalability*. The identity system should have the capacity to onboard and provide service to a significant fraction of the global population. In addition, the system should be socially scalable, have sufficient incentives for people to join the network, and have a low barrier to entry[9], including technical (private key management, interacting with specialized software, e.g. Metamask), financial (paying blockchain fees, staking value tokens or cryptocurrencies), and physical (offline proof of presence ceremonies, peer to peer physical vouching meetings.).

## 3.2 Applications

In this section, we outline existing and theorized applications of successful Proof of Personhood protocols. Each of these requires sybil-resistance to succeed, particularly at scale:

- *Universal Basic Income.* A successful program of cryptocurrency-based Universal Basic Income must ensure that recipients cannot create fake identities to receive multiple UBI allotments and defraud the system (Howitt 2019). PoP protocols serve as protection against this kind of attack, enabling the equitable distribution of cryptocurrencies to anyone who is a participant within their networks. Currently, the Idena Network, a Proof of Person blockchain reviewed in this article, is distributing mining rewards that amount to the equivalent of USD $50-60 / month*[10] for over 4000 nodes. Another example is that of the Duniter protocol, also reviewed in this article, where participation is incentivized with Universal Dividends in Ğ1 cryptocurrency, where each member receives 1 UD / day. The value of UD in Ğ1 units changes every six month as a predefined proportion of the total monetary mass of Ğ1. Currently 1 UD = 10.16 Ğ1[11].The Ğ1 economy is completely independent from other currencies which causes its value to vary with the level of adoption on different localities. Rough community estimations have placed it around $30 / month, although the theme is subject to continuous debate[12].

---

[9] 'Money, blockchains, and social scalability', Nick Szabo, 2017
[10] As of September, 2020.
[11] Duniter's Ğ1 currency dashboard, Duniter, 2020
[12] Ğ1 Forum, https://forum.monnaie-libre.fr/t/donnez-votre-estimation-du-taux-de-change-g1/1987

- *Peer-to-Peer Governance*. As discussed in Section 2, a robust mechanism for signaling unique identities is a requirement for democratic governance in peer-to-peer systems. Absent such a protocol, existing Proof of Work and Proof of Stake systems establish resource-based membership mechanisms, which result in a plutocratic model that in most cases makes voting expendable (De Filippi 2019). Thus, PoP is a requirement for meaningful governance practices within open networks.

- *Public Goods Funding*. Addressing global-scale challenges, such as climate change, pandemics, the refugee crisis, wealth inequality, etc., requires the ability to deftly coordinate across territorial or institutional boundaries to support supranational public goods. So far, most public goods have been supported by centralized entities, but this nation-state system has thus far proven largely ineffective in addressing global challenges. Traditional Web 2.0 platforms attempted to bridge this gap, facilitating voluntary individual adherence to movements (e.g., "Extinction Rebellion", "Fridays for Future") aiming to provide a response to these challenges (Bennett 2012). However, lacking reliable means to formalize unique identities, these movements are rarely able to transform the *noise* of social-media-organizing into a clear *signal* represented by votes or financial commitments. If adopted at a large scale, blockchain-based PoP protocols could contribute to a networked social infrastructure to enable this transition.

- *Quadratic Voting and Quadratic Funding*. These algorithmic mechanisms for mathematically optimal collective decision-making and resource allocation reward diversity of support more strongly than they reward the individual intensity of preference. In other words: a group of people voting or allocating resources to an option has a higher impact than a single individual expressing that same amount of support. Such designs imply that splitting one's votes or funds across multiple accounts increases the impact one is able to exert, thus creating a high incentive for Sybil Attacks. Therefore, an anti-Sybil identity system is a requirement for their application within the context of open, peer-to-peer networks (Lalley and Weyl 2012; Buterin et al. 2018). Gitcoin, a crowdfunding platform for open source projects which employs quadratic funding, is currently planning to implement two of the solutions reviewed in this article, BrightID and the Idena Network, by December 2020[13].

- *Social Media*. Social media signaling methods (impressions, likes, upvotes, etc.) have become key to societal debates, but are prone to manipulation by bots (Ferrara et al. 2016). Strong Sybil protection of social media accounts could help address the spread of fake news and fake impressions, as well as digital advertisement related frauds. Currently the Idena Network is implementing their Proof of Personhood solution for their internal forum[14] and Discord channel[15].

- *Airdrops*. A popular way to advertise new blockchain projects is to distribute ("airdrop") a fraction of cryptocurrency tokens to a wide distribution list (Harrigan et al. 2018). However, it is very common for those systems to suffer Sybil Attacks, even when requiring different forms of identification such as a Telegram account or passports. This has led to a switch toward airdrops proportional to user balance of a specific coin and lockdrops (where users need to lock some coins and receive tokens proportionally to signalling)[16]. Those techniques, despite being Sybil resistant, privilege users already having a high amount of crypto-holdings. Thus, Proof of Personhood protocols are a key component in enabling egalitarian airdrops.

---

[13] [Roadmap](), Gitcoin
[14] Idena Forum, https://forum.idena.website/
[15] Idenauth, https://github.com/iyomisc/idenauth

- *Decentralized Oracles*. Oracles are used as a mechanism to provide factual offchain data to blockchains and smart contracts (Teutsch 2017). To prevent collusion and coordinated majority attacks on oracles, in addition to Skin-in-the-Game staking, many existing protocols employ KYC for identity verification, which leads to centralization of their services. Proof of Personhood networks are able to provide an alternative solution for decentralized oracles: witnesses can be randomly sampled from the set of human participants to reach a consensus on arbitrary evidence (fact, proof of event outcome, resolution of a dispute, or even opinion poll, etc.).

- *Peer-to-peer economy*. A Sybil-resistant network utilizing the same currency can have peers directly engage with each other on transactions, exchanges and co-production of goods and services without the need of intermediaries to establish trust (Selloni 2017). Currently Duniter, one of the Proof of Personhood protocols reviewed in this article, utilizes Ğchange[17], an active internal market for exchanges denominated in their Ğ1 currency, and ğannonce[18], a crowdfunding platform for ğ1 projects.

## 4 A Taxonomy of Approaches

Before reviewing existing solutions, in this section we will outline the different theoretical primitives that underpin Proof of Personhood approaches. Throughout the past few decades, different methods have been outlined in order to address one fundamental question: how can we distinguish a human from a machine? We describe recent approaches below.

### 4.1 Reverse Turing Tests

In the opening of his 1950 paper, "Computing Machinery and Intelligence", Alan Turing asked, "Can machines think?". In order to narrow this question down to one with an objective answer, Turing created the "Imitation Game", in which an evaluator having a conversation with another entity through a text-only channel attempts to determine whether the entity in question is a human or a computer. Known as the Turing Test, to this day it is applied at an annual competition in artificial intelligence, The Loebner Prize[19], that rewards the most human-like computer programs, based on subjective assessment from human judges (previously a panel, and as of 2020 evaluated by the public).

This method created the base for a reverse test, the CAPTCHA, a "Completely Automated Public Turing test to tell Computers and Humans Apart", widely used to elicit proof from humans that they are not bots. It does so by requiring humans to parse through distorted words and images; a class of tasks known as "AI-hard" (von Ahn et al. 2003): difficult for an algorithm to perform, simple for a human. However, in addition to serving the purpose of authenticating humans, the input from a CAPTCHA test is also used to calibrate the pattern recognition capacities of artificial intelligence algorithms. Thus, machine learning presents an evolving threat to the functioning of these CAPTCHAs.

---

[16] Proof of Humanity: https://docs.google.com/document/d/1z01MS0-h75ESVmWymU2Gv3Z43p35oZAFtQLStOeu7Ek/edit?usp=sharing
[17] Gchange website, https://www.gchange.fr/#/app/market/lg?last
[18] Ğannonce website, https://gannonce.duniter.org/#/
[19] The Loebner Prize, https://www.ocf.berkeley.edu/~arihuang/academic/research/loebner.html

New approaches are being developed in order to address this challenge. The Idena Network, outlined in section 5.1, has shown that, in order for CAPTCHAs to resist the dynamic development of AI connected with neural networks and deep learning, they must not be generated algorithmically, but instead created by humans[20]. Only then will those tests move out of the class of "recognition" tasks, solvable by neural networks, and instead be classified as AI-hard problems, requiring an understanding of implied meaning, or the use of common-sense reasoning.

Recent AI-hard tests extend principles from the Winograd Schema Challenge (Levesque et al. 2012), which would pose implied-meaning questions like the one below:

*"The trophy would not fit in the brown suitcase because it was too big. What was too big?"*
*1. The trophy*
*2. The suitcase"*

However, due to its reliance on textual representation, WSCs may be vulnerable to new advances in natural language processing such as GPT-3 (the accuracy of the state-of-the-art models in WinoGrande challenge currently reaches 0.7-0.85 AUC compared to 0.94 AUC human performance[21]). Additionally, this approach requires specific language knowledge, and therefore fails to create a standard that can be applied internationally. Thus, the use of images is more likely to remain robust in the long term.

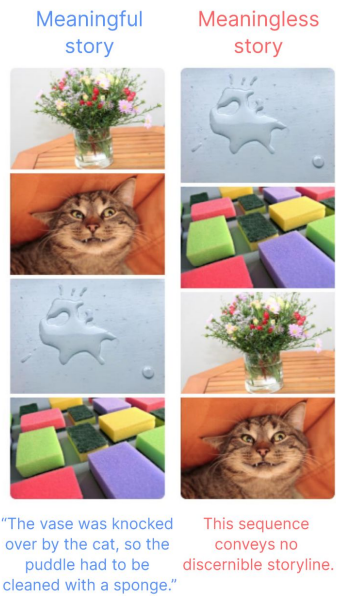

Fig.1. The FLIP test example

The Idena Network builds on these previous approaches by creating an AI-hard test that both requires common-sense reasoning, and is based on visual representation. Named FLIP[22], it asks users to choose between two orderings of images, with only one set conveying a logical and meaningful story. Human accuracy in solving FLIPs is at 95%, while AI teams have been able to reach 60-76%[23]. Alternative AI-hard reverse Turing Tests are VCR (Zellers, Bisk, Farhadi and Yejin Choi 2018), ROPES[24] (Reasoning Over Paragraph Effects in Situations), ALFRED[25] (Action Learning From Realistic Environments and Directives) and others, although to our knowledge these are not being employed by any Proof of Personhood solution at the moment.

One important note is that, while reverse Turing tests may prevent automated and bot attacks, they fail to address human-generated attacks, in which one individual passes the test multiple times and creates multiple different identities. To address this second aspect of the challenge approaches such as the FLIP created by the Idena

---

[20] 'AI-Resistant CAPTCHAs: Are they really possible?', Idena Network, 2019.
[21] WinoGrande Challenge, Allen Institute for AI.
[22] Flip Challenge, Idena Network.
[23] The "IDENA AI" with 60% accuracy at solving FLIPS', Jan Moritz, 2020.

[24] Reasoning Over Paragraph Effects in Situations, Allen Institute for AI
[25] Action Learning From Realistic Environments and Directives, Allen Institute for AI

Network, or ATUCAPTS (Andersen and Conitzer 2016) restrict participation via employing elements of a Pseudonym Party, as described below.

**4.2   Pseudonym Parties**

Pseudonym parties are an effective mechanism to avoid a tradeoff between accountability and anonymity in digital domains. This is a 'back to basics' approach that builds upon a simple security foundation: real humans can only be in one place at a time (Ford 2012). In this method, authentication comes from being physically present at a specific place and time. In this physical space, attendees will formalize procedures to register their presence, such as individuals scanning each other's QR codes, and by that act generating an anonymous credential or token (Borge et al. 2017). These credentials can then be utilized to establish membership in online communities. In essence, pseudonym parties act as a framework for individuals to conduct reverse Turing Tests on each other. For as long as hyper-realistic digital forgeries (deep fakes) remain in the Uncanny Valley (Mori 1970; MacDorman, and Kageki 2012), virtual pseudonym parties may be feasible[26]. This is especially true if utilized in combination with AI-hard tests, which require human interpretation. In these combined approaches users willing to be authenticated need to perform the test simultaneously. Thus, due to the fact that the same person cannot perform two actions at the same time, the protocol ensures the singularity of authentication.

Pseudonym parties provide significant accountability since membership rights are limited and therefore can be revoked, while preserving a relatively high degree of anonymity, since credentials do not need to contain any personally identifiable information.

However, the requirement of significant engagement for authentication is a clear downside, particularly since credentials are not permanent: all "nodes" must be synchronized at a certain frequency, so that new individuals may join the protocol. There are also concerns regarding the authentication of users in remote or faraway locations, who may not be able to attend physical gatherings. This friction may be minimized by leveraging existing gatherings such as conferences, rituals and civic ceremonies for identity authentication (Ford 2020), as well as through virtual pseudonym parties.

**4.3   Web of Trust**

A Web of Trust consists of identity certificates that can be digitally signed by other users who, by that act, declare the certificate valid, and thus provide Proof of Personhood. Through this process, the nodes of the network are effectively partitioned into Sybils and non-Sybils (Viswanath et al. 2010). The Web of Trust paradigm relies on the fact that, while an attacker may be able to arbitrarily create Sybil identities in social networks, it will be much harder to substantiate said identities with an arbitrarily large number of false connections to trusted nodes (Viswanath et al. 2012). Sybil nodes will thus be poorly connected to the trusted network, and easily identified. Web of trust schemes may be further reinforced by a reputation system that serves to track trust levels and prevent deception (Dunphy and Petitcolas 2018).

---

[26] 'Pseudonym Pairs: A foundation for proof-of-personhood in the web 3.0 jurisdiction', Johan Nygren, 2019.

We note that there have been attempts to create a web of trust framework through automated graph analysis of existing social trust networks, particularly social media networks. One such approach is SybilRank, which aims to identify fake accounts within bounded social media networks, and has met with some success (Cao et al. 2012). However, it is unlikely that widely-used online social networks are good candidates for large-scale identity approaches, particularly for sensitive applications like civic engagement. This is due not only to the ease with which attackers can create 'false' nodes with real relationships and connections to other nodes (Ford 2020; Ferrara et al., 2016), but also because re-orienting an identity program around privately-owned, centralized social network platforms is antithetical to the project of self-sovereign identity solutions.

Despite the long-term interest in Web of Trust, with the first, limited-scope version set out in the second PGP manual in 1992, several inherent issues with the approach have prevented large-scale adoption. First, a combination of different claims and credentials may not entirely guarantee sybil-resistance (Wang and De Filippi 2020). Further, levels of trust cannot be easily quantified, and only first degree relationships can be fully trusted, which can constrain the network. Similar to pseudonym parties, these issues can also prevent users from low-infrastructure or remote locations from acquiring key signs, or building in-network credibility (Wilson and Ateniese 2015). To correct for these issues, the web of trust paradigm has been adapted, extended, and paired with other approaches, notably in the form of mutual surety graphs, as well as graphs with other topological features (Shahaf et al. 2019). Two of the implementations we will discuss also aim to extend the subjectivities of the web of trust approach, moving towards a more intersectional paradigm, as described below (Immorlica et al. 2019).

## 4.4 Intersectional Identity

Intersectional Identity is a framework that aims to bridge formal verification methodologies and the informal mechanisms through which individuals check the validity of identity-related claims. It builds upon traditional Web of Trust schemes by expanding the scope of markers that can be taken into account, such as one's name, age, address, gps history, interactions, skills, work, education, etc. All of these different markers can be translated into bits, so any given individual is associated with an exponentially large number of potential bits that may be useful for authentication.

This framework achieves uniqueness, or Sybil-resistance, by drawing from three aspects of identity highlighted in the classical sociology of Georg Simmel: sociality, intersectionality and redundancy (Schützeichel 2013). Here, sociality refers to the fact that every aspect of identity is shared. Intersectionality implies that the set of others with whom the identity markers of any given individual are shared differs for each marker, thus no individual or group can serve as a central 'chokepoint' for identity verification. Redundancy denotes that the uniqueness of an individual is over-determined by the countless unique intersections of groups or sources of trust that each person finds themselves in through the course of their lives. With data architectures put in place to record intersectional markers, Sybil-resistant identities can be established by tracking just a few characteristics that uniquely identify an individual, while keeping sensitive information private (Immorlica, Jackson, and Weyl 2019).

## 4.5 Token Curated Registry

The Token Curated Registry (TCR), in contrast to the schemes outlined above, was not originally devised as a method for identity verification. In essence, TCRs draw from work on incentive systems designed to replace list owners through the creation of economic incentives for decentralized list curation. In this scheme, members of a registry hold tokens associated with the list, which may increase in value if they are able to maintain its quality, legitimacy or popularity, thus attracting more list applicants who want to add their data to it (Asgaonkar and Krishnamachari 2018). Members can establish trust through different mechanisms, such as staking a certain amount of funds, voting, or vouching for other members accurately. TCRs have successfully been applied towards curating professional profiles, media content, and other services, and are particularly instrumental in enabling decentralized courts for blockchain-based dispute resolution frameworks[27]. Building upon these successes, different identity solutions employ this mechanism to create an incentive for members of an identity registry to go through the effort of verifying each other's uniqueness and singularity.

### 4.6 Decentralized Autonomous Organizations

The DAO acronym refers to Decentralized Autonomous Organizations. DAOs are a class of smart contract[28] (Norta 2015) devised to automate the execution of organizational governance and fund allocation. In that sense, these contracts may be thought of as an automated constitution. This organizational framework emerged as a possibility due to the creation of Ethereum, a blockchain network that permits Turing-complete computations[29] (Minks 2017), leading to the growth of smart contract development. By deploying DAO contracts into the Ethereum blockchain, organizations allow their participants to pool funds (denominated in cryptocurrencies), maintain real-time control of resources, and vote on resource allocation to different projects with governance rules that are formalized, automated and enforced by the conditions encoded into the smart contract.

This type of organizational framework is employed, in different ways, by several of the Proof of Personhood protocols described in this review. In contrast to the majority of smart contracts, which serve strictly financial purposes, DAOs are highly likely to entail human decision-making in their functioning. Thus, their activities may be thought of as 'human entropy', observable on-chain, serving as a meaningful substrate for different aspects of Proof of Personhood solutions.

### 5 A Review of Existing Efforts

We now outline the approaches of these new and natively digital sources of authentication - their attributes, methodologies, strengths, and weaknesses - and sketch out possible directions for future developments.

### 5.1 Idena Network

Idena is an open source project created by an anonymous group of engineers and computer scientists. It has its own blockchain, which is driven by a proof-of-person consensus, with every node linked to a cryptoidentity with equal voting power —thus it is a fully decentralized solution. The Idena Network implements a novel way of achieving Sybil resistance by combining human-generated reverse Turing tests (Idena's FLIP test is described in detail in section 4.1 Reverse Turing Tests) with elements of a virtual pseudonym party[30].

To join the network, participants must attend live authentication ceremonies, held simultaneously for the entire network. During these synchronous events, one must complete a set of FLIP tests within a limited amount of time. Afterwards, users are required to create new tests. This is an important element, given that in order for the FLIP test to resist machine learning and truly belong to the category of "AI-hard", it was designed to not be fit for automated, algorithmic generation. Instead, FLIPs are always created by humans.

---

[27] 'Kleros: Short Paper', Clement Lasaege, Federico Ast, and William George, September 2019.
[28] 'Smart Contracts', Nick Szabo, 1994
[29] 'Ethereum Whitepaper', Vitalik Buterin, 2014
[30] 'How Idena Works', The Idena Network, 2019.

The frequency of those authentication ceremonies is determined by the size of the network —currently they are conducted around once every two weeks. Given that tests cannot be solved by existing AI, Idena successfully provides a proof of personhood. However, it is not strictly anti-Sybil, with a probabilistic margin of error: although highly unlikely, a person with exceptional ability could solve more than one set of FLIPS within their allotted time, thus earning more valid identities within the network.

As an additional layer of security, Idena requires new members to present an invite code to be able to join their first authentication ceremony. This code can only be obtained through existing members, thereby creating a Web of Trust. This also extends into a reward-based system: at every validation ceremony, Idena rewards all of its members with its $IDNA cryptocurrency; by inviting members who consistently attend validation ceremonies one may gain compounded rewards.

Launched in August 2019, to date the Idena Network has been able to validate 4012 identities[31]. Their approach presents a significant advance for the research and development concerned with natively digital identity protocols. This network demonstrates that combining human-generated AI-hard tests with "liveness" — a synchronous event— can play a critical role in Sybil prevention: the time constraint prevents a single entity from solving more than one set of FLIPs, while the human-generation aspect provides a defense against machine learning. Furthermore, the protocol protects privacy as it involves no data point except that of proof of conscious cognitive ability.

The most salient tradeoff in this system is the significantly high coordination cost to achieve recurrent, simultaneous solving of FLIPS: all nodes must continuously participate in the synchronous events, otherwise their identities expire. This reduces the incentive for nodes to join the network, although the relative value of the rewards paid by the protocol for successful validation and participation in block producing, currently at ~$50-60 USD / month, may succeed in creating a new habit for users. Additionally, it remains to be seen whether Idena's Sybil-resistance strategy will be able to weather the dynamic development of AI connected with neural networks and deep learning. Furthermore, the robustness and long term effectiveness of their incentive systems may also be tested in the future by the creation of markets that sell false identities and/or attacks by mechanical turks.

## 5.2 HumanityDAO

Humanity DAO is an Ethereum-based protocol. It was designed to incentivize a set of economic actors to maintain a registry of unique human identities without a central authority[32], and leveraged existing work on Token Curated Registries (Asgaonkar and Krishnamachari 2018). In Humanity DAO's case, holders evaluate candidate identities and deem them legitimate through consensus-based voting. The protocol consists of the following steps:

1. Applicants made a request to join the list using their social media profile information.
2. Applicants staked a fee on their candidacy. If the applicant got rejected, the application fee was ceded.

---

[31] 'Epoch #0054', Idena Network, 2020
[32] 'Introducing HumanityDAO', Rich McAteer, 2019.

3. Members of the list voted on whether the new applicant should be included based on the submitted profile. Members were incentivized to curate the list honestly in order to generate demand from new applicants, leading to a long term sustainability of the project.

The registry had a method called *isHuman* that any smart contract could query to see whether a given Ethereum address had been confirmed as a unique human. Humanity DAO also deployed a Universal Basic Income smart contract with 2,500 Dai (~$2,500), which early applicants could claim at a rate of 1 Dai per month, until supply ran out (Chen and Ko 2019).

Launched in May 2019, the project quickly gained rapid traction, reaching around 640 approved members and being adopted by many influential figures within the Ethereum Network, however growth stagnated after the initial community of early adopters from the network was saturated[33]. Further, as a fully decentralized solution, creators had very little ability to change the protocol after it was launched. As related to us by the founder, this resulted in Humanity DAO suffering various forms of attacks, including one in which a change to the smart contract made it prohibitively expensive for new applicants to join. These repeated attacks led to the eventual termination of the project in January 2020.

## 5.3 Kleros

Kleros is an Ethereum-based protocol for decentralized dispute resolution. Their successful experiments with TCRs for distributed courts led the team to propose "Proof of Humanity"[34]: a solution for identity based on TCRs combined with a web of trust and based on submitted photos, bios, and video recordings. This information will be stored using the IPFS (InterPlanetary File System). Kleros' approach distinguishes itself by appending to the functioning of its protocol a recourse to adjudicate cases of faulty or duplicate users. This is done through distributed dispute resolution systems such as the Kleros Court[35] (or if decided by members through the registry's internal governance, other alternatives such as Aragon's courts[36]).

Within the Proof of Humanity protocol, users can vouch for each other with a certain amount of financial stake. To incentivize the maintenance of the registry, vouching deposits serve as a bounty, available for anyone able to correctly identify false positives in the registry. If a member vouches for users that are later determined to be duplicate or false by the distributed court, they are punished in the form of being removed from the registry and losing their vouching deposit, thus discouraging such attacks[37].

While this protocol has significant promise in building an effective reputation-based web of trust with tools in place to adjudicate cases in which the singularity of an identity is disputed, it compromises biometric information of members by requiring a video selfie and other additional information, which may de-incentivize potential users.

---

[33] 'Humanity DAO Post-Mortem', Kacper Wikiel, 2019.
[34] Proof of Humanity: https://docs.google.com/document/d/1z01MS0-h75ESVmWymU2Gv3Z43p35oZAFtQLStOeu7Ek/edit?usp=sharing
[35] Kleros Court: https://blog.kleros.io/kleros-court-revitalised/
[36] Aragon Court, https://anj.aragon.org/
[37] 'Kleros: Short Paper', Clement Lasaege, Federico Ast, and William George, September 2019.

### 5.4 Upala

Upala is an Ethereum-based protocol, designed to be interoperable with DAOs[38]. It's social graph consists of verification groups that assign a score for each member, denominated in currency; this gives members the right to steal from the shared pool of the group they belong to, the amount of their score. The act of stealing (a "bot explosion"in Upala terms) automatically deletes their identity.

Thus, this model implements the social responsibility concept, in which groups are incentivized to develop approval mechanisms that lead towards having highly-trusted members. Any existing DAO may fit into the Upala protocol, given that members are willing to trust each other by collateralizing funds in exchange for distributing reputation. Groups can be also be composed by direct end users or other groups —thus combining uniqueness scores into larger pools.This framework generates a market for identity authentication where on the supply side groups are trying to gather as many users as possible (through subgroups or directly), with the highest reputation (i.e. lowest risk of explosion) and the maximum deposits; and on the demand side, users are trying to get the highest scores for the lowest investment of reputation or money.

The Upala model expands on the principles behind a Token Curated Registry (where members are incentivized to maintain a high quality list) and also employs an intersectional lens by enabling different schemes to be created and combined within its protocol.

Given that the uniqueness scores are, to a certain extent, relative to pooled funds, this may lead to capital-rich users having an ease in obtaining higher scores—although groups may establish different verification mechanisms capable of placating this vulnerability. Another major vulnerability encoded within this model would be an avalanche user exit: if an event leads to loss of trust in Upala, an avalanche of individuals may explode their identities in panic to seize assets, ignoring the reputation consequences. However, it is possible that the probability of such a scenario materializing decreases as trust consolidates within the system, with increased usage by third parties for scoring users.

Upala has launched its first working prototype on the Kovan testnet of Ethereum in June 2020.

### 5.5 BrightID

BrightID operates an intersectional web of trust protocol, built through graphing social connections, with the additional input of trusted seed identities. The purpose of this protocol is to allow for users to provide proof that they are not using multiple accounts on a single application, and it is thus designed to be interoperable with Web 2.0. social media platforms. The interconnectivity of its graph is designed to identify true identities and Sybil identities, based on node position in relation to trusted seeds.

BrightID is the solution most in line with the Intersectional Identity paradigm, formalizing social connections in order to allow for a variety of nodes to join the system and customize their own evaluation criteria. In that sense, there are no obvious limits to the number of trusted seeds in the BrightID graph: any application utilizing their authentication solution may establish its own BrightID node with different trusted seeds. Each BrightID node runs its own instance of ArangoDB to store the graph of Web of Trust connections. Every verification can be broadcasted to a specified isolated smart contract on Ethereum or another blockchain.

---

[38] Upala's Documentation: https://upala-docs.readthedocs.io/en/latest/

The social graph serves as a common base across all nodes, but the analysis of that same graph can be distinct, so the protocol does not require consensus across nodes. Applications may either run their nodes in a centralized or closed manner, sharing their analysis and verification outputs only with themselves, or they can provide a greater level of decentralization, allowing any user to run the verification and sample the output from a large number of nodes. To control for Sybil attacks BrightID runs GroupSybilRank, a modification of the aforementioned SybilRank algorithm, to estimate the anti-Sybil score of the network participants based on affinity between groups. Proposed to be used as the official BrightID anti-sybil algorithm[39], the effectiveness of this algorithm in the presence of multiple attack vectors, remains to be proved.

BrightID's open Web of Trust architecture is robust and promising. That said, at this early stage BrightID's social experiment has significant challenges to overcome in terms of Sybil-resistance, decentralization, self-sovereignty and privacy. As of July 2020, its solution is limited to a small seed network, so there are no established paths for individuals or groups who are completely independent from the existing network to self-authenticate - thus it is not yet a fully self sovereign solution. This is not an intractable limitation, as new nodes can potentially define new verification methods that would allow for islands of users to be verified. However, scaling this process is far from trivial. One possible solution is through establishing partnerships with existing social media platforms that reach a wide net of users, but this would largely defeat the aim of the initial motivators of Proof of Personhood solutions. Therefore, the crucial challenge for the success of this experiment is finding a path forward for scalability while maintaining decentralization.

To this end, BrightID's whitepaper encourages the creation of new seeding DAOs (Decentralized Autonomous Organizations), and establishes that the BrightID Main DAO will promote research of different seed selection methods, as well as the creation of tools that can make seed selection scalable. In that sense, BrightID's success may be in tandem with an increase in the adoption of decentralized governance frameworks. Another possible pathway to scale is through BrightID's weekly pseudonym parties, during which prospective members can meet the existing community and form new links to obtain verification.

A new blockchain, IDChain, was recently introduced to implement BrightID DAO governance[40]. IDChain is a fork of the geth Ethereum node software, at which BrightID participants can self register via a web-service to receive a lifetime supply of Eidi (the native gas token on IDChain). Hedge for Humanity, a U.S. based, tax deductible 501(c)(3) charitable organization, plans to start distributing $1 US dollar / month to each of the BrightID's users as a Universal Basic Income, as a way to incentivize attacks that can provide greater visibility into the vulnerabilities of the identity system. Currently BrightID has 556 users with a positive anti-Sybil rank[41].

## 5.6 Duniter

The Duniter project, originally named uCoin, was started in June 2013[42]. The project is a technical implementation of the relative theory of money (RTM), developed by Stéphane Laborde, where a

---

[39] 'BrightID Anti-Sybil', BrightID, 2020.
[40] 'Introducing ID Chain', Adam Stallard, 2020.
[41] BrightID data, BrightID, September, 2020
[42] Duniter License, https://duniter.org/en/wiki/g1-license/

Universal Dividend is described as having its value relative to its monetary mass. Duniter is an independent blockchain utilized to mint the Ğ1 cryptocurrency as a Universal Dividend available to unique human participants. Authentication within the Duniter protocol is done through a Web of Trust type of scheme.

In order to become a part of Duniter's Web of Trust, one must receive five different vouches from existing members[43]. Duniter members are required to check a statement where they agree to vouch solely for new applicants who they have met in the physical world, or know enough to be able to contact remotely through different channels, such as social network, forums, email, video conferences, and phone calls[44]. For each new member, a cryptographic key pair is created. Furthermore, any newcomer must be at a maximum distance of 5 different connections from "referent members", which can be thought of as more central and highly trusted seed identities. Referent members are defined by a graph property that intends to mimic trust parameters of real world, human, social graphs[45]. These requirements imply that it entails significant time for new certifications to be emitted, allowing the network to manually monitor and mitigate attacks. By doing so, the Duniter protocol delegates technical governance to human governance, while still maintaining a relatively high degree of decentralization. This Web of Trust is seeing a slow but steady growth in France and nearby countries. As of September 2020, the Ğ1 currency had 2801 holders[46].

## 5.7 Equality Protocol

The Equality Protocol approach creates a meta protocol against which other identity protocols can measure their legitimacy[47]. It is designed to create an intersubjective space able to account for measurements of both collective intentionality and objective facts. It does so by combining a subjective function that provides legitimacy to the score based on Quadratic Voting, and an objective function that measures the Gini Coefficient of any DAO existing on the Ethereum blockchain. It will create a Democratic Index, as shown in the figures below, and assign a score to every Ethereum address relative to the intersection of DAOs in which it belongs as a member, or its position in the social graph of blockchain-based transactions.

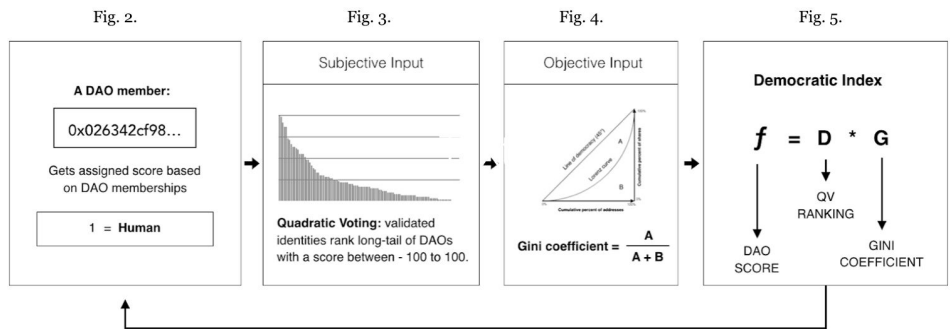

Fig. 2. Every address that belongs to a DAO will be weighted by the Intersubjective Consensus protocol with a percentage of the Democratic Index obtained from the intersection of DAOs that constitute the identity of such address.

---

[43] Web of Trust settings, Duniter, 2020
[44] Duniter Forum, https://forum.duniter.org/t/nombre-didentites-dexclusions-de-revocations-etc/7428
[45] 'Deep-dive into the Web of Trust', Duniter, 2018.
[46] Duniter's Ğ1 currency dashboard, Duniter, 2020.
[47] 'Equality Protocol', DemocracyEarth, 2020.

Fig. 3. In order to counterweight false positives on the Gini Coefficient, addresses that achieve a high score will be granted the right to rank the different DAOs analyzed by the Intersubjective Consensus oracle, according the their corresponding ability to ensure that no single Human controls more than one identifier within its domain.

Fig. 4. The Gini coefficient for democracy ranges from 0 to 1, with 0 representing perfect totalitarianism and 1 representing perfect democracy. It computes a score that measures the share distribution by each segment of addresses belonging to a DAO.

Fig. 5. A Democratic Index is calculated for each DAO, as a function of their position in the Quadratic Voting rank and their Gini Coefficient.

The current interface provides a basic personhood calculation for each DAO member of v1 Moloch DAOs, with 529 addresses receiving a probabilistic human score [48]. Examples of DAO contracts with parameters applicable as inputs for the Equality Protocol's Probabilistic Anti Sybil Score oracle include MolochDAO, DAOstack, Kleros and Aragon DAOs. Additional sources of trust can be included: different credentialing mechanisms could instantiate their protocols through a DAO, while other non-EVM protocols such as the Idena Network could be made compatible by allowing token swaps.

This approach intends only to be a meta-protocol contemplating existing DAO members, rather than forming a Proof of Personhood substrate in and of itself. However, in addition to requiring that new applicants get voted in, DAOs are usually joined through the staking of resources. Thus, the system may favor capital-rich users, who could find their way into several DAOs with more ease. This is not an intractable challenge, as it is possible to earn DAO shares through non-financial contributions and different membership protocols could potentially make the environment more inclusive over time. Furthermore, while the Quadratic Voting function brings a desirable governance component to identity verification, it remains to be seen whether it can serve as a rigorous substrate to signal the legitimacy of DAOs, given the inherent conflicts of interest that emerge due to the impact of results on the probabilistic score of voters.

## 6 Discussion

The seven reviewed projects in the digital identity space have made previously unimaginable progress in creating robust, repeatable paradigms to construct a PoP solution. They approach the problem in a diversity of ways, and use a variety of different substrates in order to successfully authenticate human users: reverse Turing Tests (Idena Network), social graph data emerging from a Web of Trust type of scheme (Duniter, BrightID, Humanity DAO and Kleros), and intersectional approaches that combine an analysis of objective financial value and its distribution within different domains, with some other form of human entropy or social signaling that can be detected online (Equality Protocol and Upala).

Despite clear shortcomings present in each of these methods, their creative uses of subjectivity can point towards interesting, hybrid approaches to verifying Proof of Personhood. In fact, the distinction between methods for Sybil-protection may be overdrawn: most of the solutions outlined in this review employ a combination of them in order to secure their networks. In some cases this combination of tactics is formalized, such as with Idena Network - predicated on reverse Turing tests, but accessible through their invite code system, an instance of a Web of Trust approach. In other cases additional protocols are appended tangentially and informally, such as with BrightID's employment of weekly Pseudonym Parties to welcome new members. In this sense, the theoretical primitives have almost false separations: when it comes to implementation, they truly co-occur and build on one another, rather than being contained by the distinctions we see explicated in the academy.

---

[48] 'Sovereign Dapp', DemocracyEarth, 2019.

Below is a comparison of the different properties of PoP solutions as established in the literature. We consider each approach in light of not only its base-level attributes - governance structure, size, blockchain, protocol, and substrate, but also its performance on the primary desired properties of decentralization, privacy, and scalability.

**Table 1: Approach Characteristics and Comparisons**

| Approach | Governance | Size | Blockchain | Protocol | Substrate |
|---|---|---|---|---|---|
| Idena Network | Public network | 4012 | Idena | Synchronous reverse Turing-test ceremonies | FLIPS |
| Humanity DAO | DAO | 640 | Ethereum | Web of Trust + Token Curated Registry | Social media information |
| Kleros | DAO + Legal entity | - | Ethereum | Web of Trust + Token Curated Registry | Personal information, photos, and video selfies |
| Upala | DAO | - | Ethereum | Price of Forgery | Identity + controlled proof of stake |
| BrightID | DAO | 556 | Ethereum/ IDChain | Intersectional Web of Trust | Existing social connections + weekly online meetups |
| Duniter | Public network | 2801 | Duniter | Web of Trust | Contact information + 5 connections to existing members |
| Equality Protocol | DAO | 529 | Ethereum | Meta Protocol: Democracy Index | Participation in DAOs |

**Table 2 : Technical Analysis**

| Approach | Decentralization | Privacy | Scalability |
|---|---|---|---|
| Idena Network | Decentralized identity registry management. Every participant can run a full validator/mining node. Change of the protocol requires network consensus | No personally identifying information (PII) sharing required. Node IP address observable | Node install, getting invite code and regular participation in the validation ceremonies required. Participation incentivized with mining and ceremony rewards (~$1.5-2/day) |
| Humanity DAO | P2P vouching. Decentralized identity registry management | Social media account (Twitter) sharing required. Ethereum address observable | Web3 dapp interaction, paying Ethereum fees and identity stake required. Participation incentivized with a UBI (1 Dai/month) |
| Kleros | P2P vouching. Decentralized identity registry management | Video selfie sharing required. Ethereum address observable | Web3 dapp interaction, paying Ethereum fees and identity stake required. Participation incentivized indirectly with court rewards |
| Upala | Decentralized identity registry management | No additional PII sharing required. Ethereum address observable | Web3 dapp interaction, paying Ethereum fees and identity stake required |
| BrightID | P2P vouching. Semi-decentralized identity registry management | No PII sharing required. Social graph and IDChain address observable | Mobile app vouching required. Participation incentivized with IDChain native token airdrop that can be used to pay fees |
| Duniter | P2P vouching. Decentralized identity registry management | No PII sharing required. Social graph observable | Vouching meeting with 5 members required. Participation incentivized with Universal Dividends in Ğ1 cryptocurrency (~$1/day) |
| Equality Protocol | Decentralized identity registry management | No PII sharing required. Ethereum address observable | Web3 dapp interaction and paying Ethereum fees required |

As outlined in Tables 1 and 2, each project entails significant tradeoffs, with each substrate forming or leading to a possible weakness in the system. Here, we outline the research gaps evident in each protocol, with the hope of providing a path forward to addressing and solving these issues.

We begin with the Idena Network, the only fully decentralized and privacy preserving solution. Currently, the synchronous reverse Turing test model of the network requires a significant commitment of time and effort on the part of its participants, who must participate in regular validation ceremonies approximately every two weeks. While its Sybil-resistance strategy is currently effective, it remains to be seen whether AI-hard tests will be able to resist the dynamic development of AI connected with neural networks and deep learning. Furthermore, it is not certain that the current incentive model put in place will be sufficient to disincentivize the creation of a marketplace for false identities with mechanical turk attacks.

Humanity DAO, while extremely promising, required the use of privately-owned identity information from social networks like Twitter to verify identity, again exposing users to the vulnerabilities of Internet monopolies and largely defeating the aim of the initial motivators of such consensus identity proofs. The system also fell prey to attack due to its necessarily fixed protocol. Kleros requires users submit a range of personal information and video proof - effectively a biometrics, which is likely to prevent many from using the service, and it remains to be seen whether their system of reward and punishment will be sufficient to prevent dishonest vouching. Upala's social responsibility concept shows promise in preserving trust, but may be more accessible to capital-rich users, given that uniqueness scores are in part relative to stake - although this may be contemplated by different authentication methodologies or governance rules established by groups adopting the protocol. Upala's protocol also runs the risk of suffering an "explosion" avalanche, with users exiting the protocol en masse.

Duniter's requirement of at least five vouching links, and a maximum distance of 5 different connections from referent members, imply that it entails significant time for new certifications to be emitted. This deliberately slower rate exhibits good Sybil-protection properties, but significantly restricts network growth. Currently, this is not particularly problematic, as there is not yet a plurality of implementations that would allow for the creation of a proper benchmark for "adoption rate".

What is already possible to infer, from the set of projects reviewed here, is that the adoption of subjective Proof of Personhood protocols is distinct from that of other authentication technologies since they grow more slowly, at a "human rate". However Duniter's model is also restricted to its community geographics, which limits its application to local use cases, as opposed to those intended for a global audience.

The Equality Protocol, an intersubjective consensus protocol to evaluate other protocols, does not form a substrate of identity verification in and of itself and it is currently fairly restrictive in its scope, as it solely contemplates members of decentralized autonomous organizations. While DAO participation rights can be earned through non-financial means, this system is particularly vulnerable to a concentration of power with capital-rich users who can purchase the ability to join and participate in these organizations with more ease. Furthermore, while the use of Quadratic Voting is clearly appropriate as a consensus mechanism, it has not yet been fully substantiated as a method of signaling the legitimacy of DAOs. The key weakness concerns the possibility of conflict of interest: the protocol takes as ground truth user rating of DAOs according to their sybil resistance; however, there does not yet exist a verifiable mechanism within the protocol to confirm whether this *perception* of Sybil resistance from the user is true or not. This may lead to misaligned incentives and inaccurate assessments.

Finally, BrightID is currently the most intersectional solution explored in this paper, and thus may have significant scalability potential. However, its current reliance on establishing trust through connections to a small, trusted seed network makes it difficult for independent groups to self-authenticate. BrightID has a certain degree of centralization, as it relies on privately configured nodes to manage identity registries, selected by BrightID founding team, although there is promising potential for improvement with the introduction of the IDChain and integration of the seed selection and vouching process into IDChain-based DAO. One more possible hindrance to the adoption of BrightID stands in its reliance on a public social graph, which may compromise the privacy of authenticated users, if the real world identity of some of the participants is revealed. Finally, the Sybil-resistance of BrightID's GroupSybilRank algorithm has yet to be proven.

As four out of the seven solutions analyzed in this review rely primarily on a Web of Trust, it is important to note that presently there is no evidence of Web of Trust schemes' effectiveness for Sybil-resistance in the presence of multiple attack vectors. Bad actors may forge multiple real relationships under different names in different groups: if there are enough non-intersecting small groups an attacker may be able to grow a significant amount of Sybils over time. The prevention of such attacks often requires sophisticated data processing and modeling techniques: a notable example is Facebook's periodic take down of, on average, two billion fake accounts per quarter using machine learning algorithms like SybilEdge, which employ behavioral and content classifiers to flag an account as abusive (Adam Breuer 2020).

Thus, we see that there is still significant work to be done. One possible mode of inquiry is to look to PoP systems that are not directly blockchain-based, but instead use more intersectional approaches. Described in Section 4.4, a theoretical approach to such a project was outlined by Nicole Immorlica et. al, proposing a protocol of verifying identity through proofs of social intersection, extending the Web of Trust approach (Immorlica, Jackson, and Weyl 2019). This system would allow for users to check the claims of others, with varying levels of trust, or credit, assigned to each user in relation to others; this system of credit could also extend to groups of users, as relevant, to further prevent false claims. Such a system has been partially implemented by Identiq, which has created a providerless, peer-to-peer network that allows for companies to collaborate to validate users[49]. However, Identiq is not only itself privately-owned and closed-source, it also puts validation power in the hands of corporations, and thus does not provide a fully decentralized solution, particularly one that could be leveraged for civic engagement purposes.

Protocols that focus directly on social interaction are also relevant here. Consider Nomqa[50], an upcoming solution that verifies humanity by scoring interactions between users based on subjective meaning. This approach brings in the much-needed subjectivity component to identity solutions, considering collective, rather than purely individual, approaches to identity. Markedly more offline solutions have also been proposed through the use of 'pseudonym parties': Personhood.online, a project developed at the École Polytechnique Fédérale de Lausanne[51], integrated physical gatherings with a DID architecture[52] and next generation blockchain technologies aimed at scalability and privacy. However this ambitious effort has been inactive since 2018. Another insightful proposal was to produce a temporary proof of personhood based on physical attendance, forming a 'seed set' comprised of said attendees[53]. These seeds can then validate other identities, creating trusted clusters, which can fan out and validate larger and larger sets and communities.

Additional possible directions of future inquiry include explorations of blind research into social networks - expanding the possibilities of establishing trust between nodes while maintaining their privacy[54] as well as anticollusion systems[55]. A prominent use case for Proof of Personhood solutions is in the context of blockchain-based voting. However, by generating a record of transactions, blockchains can facilitate bribery, with smart contracts created to reward users if they are able to demonstrate a certain voting pattern through a publicly verifiable transaction[56]. Minimal Anti-collusion Infrastructure is a scheme, currently employed by BrightID[57], that aims to address these types of attacks by allowing voters to switch their voting keys at any time: thus one may provide a voting receipt, but can never guarantee that said vote had not been formerly invalidated by a key switch. While there are still possible vectors of attack (one could sell their private key), the Minimal Anti-collusion Infrastructure outlines a promising approach to address on-chain privacy for identities being used in voting mechanisms. Another proposal to "make honesty the optimal strategy" is to have each edge within a network acting as a prediction market: in case the legitimacy of a node is challenged, reputation flows from the losers to the winners[58]. This proposal is in line with Klero's approach of appending a distributed dispute resolution system into its protocol and using vouching stakes as a bounty available for network policing.

Any endeavor to create functioning digital democracies can be undermined by exploitation of identity, from the automated creation of false identities to corruption by third parties controlling a voter registry. Determining who has the right to participate cannot be an afterthought of democracy: it is its elemental task. However, it must also be noted that democratic governance is possible even with bounded sybil penetration, meaning that a small amount of error within a system can be forgiven, which opens up possibilities for more intersectional and subjective approaches (Shahaf, Shapiro, and Talmon 2019).

---

[49] Website: identiq.com
[50] Website: nomqa.com
[51] Website: personhood.online
[52] 'Decentralized Identifiers', World Wide Web Consortium
[53] ['Proposal for a Decentralized Unique Identity Seeding Protocol'](), Howitt, 2020.
[54] ['Private Social Network Search'](), Barry Whitehat, Kobi Gurkan, 2020.
[55] ['Minimal Anti-collusion Infrastructure'](), Vitalik Buterin, 2019.
[56] ['On-chain vote buying'](), Philip Daian, Tyler Kell, Ian Miers, Ari Juels, 2018.
[57] ['Anonymous Participation using BrightID'](), Adam Stallard, 2020.
[58] ['Towards Proof-of-Person'](), Peter Watts, 2018

Finally, the steady advancement of machine learning and artificial intelligence makes the question of formalizing identity frameworks particularly urgent. Trustworthy and high-quality information is the foundation of a functioning democracy - and yet, from deep fakes to language model outputs, machine-generated information is becoming easier to create and spread. In the future, there may be a need for cryptographic signatures on selected media or information pieces, to establish trust and authenticity (Ford 2020).

Thus, in many senses, governance, democracy, and identity are strictly correlated. Structuring communication architectures anchored on decentralized, privacy preserving, self sovereign and Sybil-resistant identity protocols that can reach all humans with an Internet connection can open the path for new, radically participative peer-to-peer political movements and economies.

## 7   Conclusions

Identity is one of our most fundamental human attributes. However, in the age of surveillance capitalism, identity itself has become a part of a new, digital political frontier[59] (Zuboff 2019). As Edward Snowden, one of the most prominent activists for the end of surveillance practices in the world, recently warned during a videoconference at the 2019 Web3 Summit in Berlin[60]: *"The one vulnerability being exploited across all systems is Identity."*

If the "State is the monopoly on violence" as Max Weber once defined it (Weber 1919), then the Surveillance State (or Surveillance Capital) is the monopoly on identity. Consolidated credential mechanisms today all verify humans by implementing practices that require the disclosure of personal and private information to an identifier. Eventually, this wealth of information accrues into credential monopolies, which are a prominent force in the perilous drift toward democratic deconsolidation now threatening Western democracies. While there is significant space for action in advancing effective public policies that contemplate those threats, approving and enforcing them is often extremely challenging in the face of the powerful market forces they stand against. In that sense, the alternative technological paradigms that may arise from Proof of Personhood systems could provide a relevant path towards guaranteeing privacy and participation rights.

Further, surveillance capitalism bears a worldview that downgrades human value and dignity in favor of machine learning systems. Proof of Personhood systems counter that logic by creating the building blocks of a human-centered economy, where individuals directly control and have governance rights over the networks, communities, and organizations they belong to. These systems invert the current logic of capitalism, creating the base for solidarity economies that can safeguard and elevate the role of human consciousness, choice, and agency.

Yes, the approaches explored in this review fall short of this goal in several ways, some still relying on existing sources of centralized information, others on small networks or high-friction synchronous tasks. Nonetheless, Proof of Personhood projects present one of the few viable alternatives capable of addressing these problems at their root. In doing so, they illustrate that the best technologies do not abstract away subjectivity. Instead, they embrace it, seeing subjectivity for what it is: not just a necessity, but a strength.

---

[59] '[The Social Smart Contract](#)', Democracy Earth, 2017
[60] [Edward Snowden](#), Web3 Summit, 2019

## 8 Acknowledgments

We would like to express our sincere gratitude to Adam Stallard, Clément Lesaege, Peter Porobov, Rich McAteer, and Vinay Taylor for sharing their work with us and contributing with invaluable feedback to this review.